\documentclass[11pt]{article}
\usepackage{graphicx}
\usepackage{subfigure}
\usepackage{xspace}

\newcommand{\BABARPubYear}    {06}

\newcommand{\BABARProcNumber} {156}
\newcommand{\SLACPubNumber} {12080}

\input babarsym

\def\Pep{\relax\ifmmode{\rm e^+}%
  \else${\rm e^+}$\fi}%
\def\Pem{\relax\ifmmode{\rm e^-}%
  \else${\rm e^-}$\fi}%
\def\PgUc{\relax\ifmmode{\rm \Upsilon(4S)}%
  \else${\rm \Upsilon(4S)}$\fi}%
\def\PKzL{\relax\ifmmode{\rm K^0_{\rm L}}%
  \else${\rm K^0_{\rm L}}$\fi}%

\newcommand{\fig}{Fig.\xspace}

\newcommand{\V}{\ensuremath{\mbox{\,V}}\xspace}

\newcommand{\microA}{\ensuremath{\,\mu\mbox{A}}\xspace}
\newcommand{\nA}{\ensuremath{\mbox{\,nA}}\xspace}
\newcommand{\nF}{\ensuremath{\mbox{\,nF}}\xspace}
\newcommand{\GeV}{\gev}

\long\def\inst#1{\par\nobreak\kern 4pt\nobreak
    {\it #1}\par\vskip 10pt plus 3pt minus 3pt}

\begin{document}
{\pagestyle{empty}

\begin{flushright}
SLAC-PUB-\SLACPubNumber \\
\babar-PROC-\BABARPubYear/\BABARProcNumber \\
September, 2006 \\
\end{flushright}

\par\vskip 2cm

\begin{center}
\Large \bf The \babar\ Muon System Upgrade
\end{center}
\bigskip

\begin{center}
\large 
W. Menges\\
Department of Physics, Queen Mary, University of London \\
Mile End Road, London, E1 4NS, UK \\
(from the \lbabar\ Collaboration)
\end{center}
\bigskip \bigskip

\begin{center}
\large \bf Abstract
\end{center}

Resistive Plate Chambers (RPCs) were used for the instrumentation of
the iron flux return of the \babar detector as a muon system. Unfortunately
the efficiency of the original RPCs degraded rapidly with time. Limited 
Streamer Tubes (LSTs) were chosen for an upgrade of the barrel portion of
the detector.

The phased installation started in summer 2004 with replacing two
sextants of the \babar barrel muon system with LSTs. The modules for
the remaining four sextants are under long-term test at SLAC and ready
for installation, expected in 2006. The modules become inaccessible
once installed in \babar, so it is critical to select only perfectly
working tubes. To accomplish this a strong QC system was established
during the prototype phase, and fully implemented throughout
pre-production and construction. To spot any damage during transport,
the final modules are subjected to comprehensive tests at SLAC
immediately after arrival and kept under long-term test till
installation into \babar. Details of these tests and results from
long-term testing will be discussed. Since spring 2005 the PEP-II
accelerator is running and \babar is collecting data. First experience
from data taking with the LSTs will be presented and the performance
of the detector discussed.

\vfill
\begin{center}
Contributed to the Proceedings of the\\ 2005 Nuclear Science Symposium
and Medical Imaging Conference, \\
10/23/2005---10/29/2005, Wyndham El Conquistador Resort, Puerto Rico
\end{center}

\vspace{1.0cm}
\begin{center}
{\em Stanford Linear Accelerator Center, Stanford University, 
Stanford, CA 94309} \\ \vspace{0.1cm}\hrule\vspace{0.1cm}
Work supported in part by Department of Energy contract DE-AC02-76SF00515.
\end{center}

\section{Introduction}

Resistive Plate Chambers (RPCs) were used as the initial
technology for the Instrumentation of the Flux Return (IFR) of the
\babar detector~\cite{Aubert:2001tu} for the identification of muons
and \PKzL.  Unfortunately, there were many issues with the production
and operation of the RPCs, which led to significant decrease in the
per-layer efficiency in the first years. For more details see
\cite{Piccolo:2001zy,Anulli:2003bk}. The \babar collaboration decided
to upgrade the forward system with improved RPCs~\cite{Anulli:2004zu}
and to replace the RPCs in the barrel system with Limited Streamer
Tubes (LSTs). The research and design phase started in 2002 with the
first installation phase in summer 2004 and the second scheduled for
autumn 2006.

The geometry of the muon system is shown in \fig~\ref{fig:ifr_geo}. 
Forward, backward and barrel system are sextant shaped. In the initial
layout the barrel system consists of 19 layers, all equipped with
RPCs. Layer 19 is not accessible and no LSTs will be installed there.
The inner 18 layers will be used for the LSTs installation. To
compensate for the lose of absorbing material between layers 18 and 19,
brass will be installed in 6 layers. The optimal location is every
second layer starting with layer 5. The remaining 12 layer will be
filled with LSTs.

\begin{figure}
  \centering
  \includegraphics[width=0.9\textwidth]{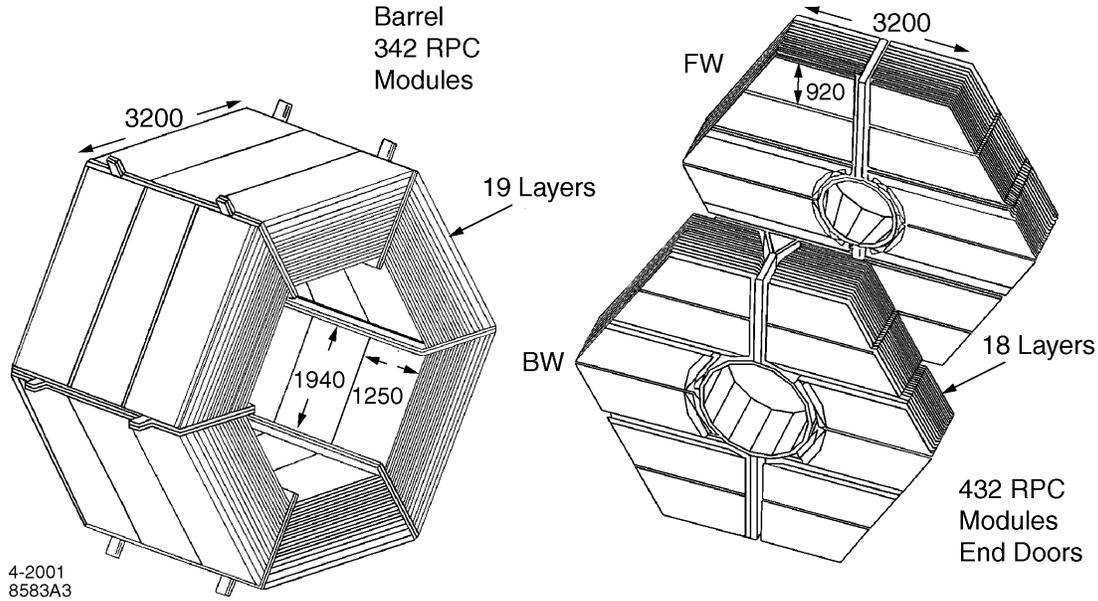}
\caption{Overview of the IFR: Barrel sectors and
forward (FW) and backward (BW) end doors; the shape of the RPC
modules and their dimensions are indicated.}
  \label{fig:ifr_geo}
\end{figure}

\section{Detector Components}

The physical principle of LSTs is quite simple. Each gas-filled cell 
has a single wire at High Voltage (HV). If a
charged particle passed through the cell, the gas is ionised and a
streamer builds up, which can be readout from the wire.
Simultaneously a signal will be induced on a plane, which is
mounted below the tube.  The charge on the wire is used as the $\phi$
coordinate and the induced charge on the plane is detected using
strips perpendicular to the wire direction, giving a z-coordinate. The
r-coordinate is taken from the layer information. Together this gives
a 3d information of the hit.

Therefore the basic components are limited streamer tubes and
z-planes. Other components are a gas and HV system and the readout
electronics. 

\subsection{Limited Streamer Tubes}

The cross section of a tube is shown in \fig~ref{fig:tube:sketch}. It
consists of 7 or 8 cells. Each cell is 17\mm wide, 15\mm high and
380\mm long. In the middle of each cell, a gold-plated anode wire is
clamped. Six wire holders are equally distributed over the length of a
cell to prevent the wire from sagging and touching the PVC walls,
which are painted with a water-based graphite paint and kept at ground
potential. Both endcaps are equipped with gas connections. One endcap
also hosts the HV connectors (\fig~\ref{fig:tube:photo}). Two wires
are bundled into one HV channel. The streamer signal is readout via a
capacitor on the HV connection.

\begin{figure}
  \centerline{%
    \subfigure[Sketch.]{\includegraphics[clip,height=3.5in,angle=90]{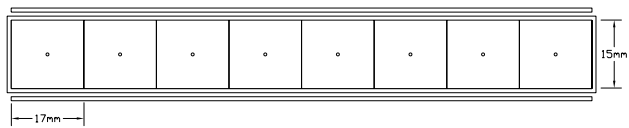}
      \label{fig:tube:sketch}}
  }
  \centerline{%
    \subfigure[Endcap with HV and gas connectors.]{\includegraphics[width=3.5in]{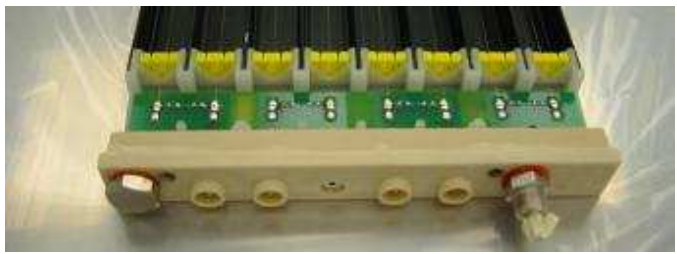}
      \label{fig:tube:photo}}%
  }
  \caption{The cross section (a) and the HV endcap (b) of a tube is shown.}
  \label{fig:tube}
\end{figure}

\subsection{Quality Control of LSTs}

The experiences with the RPCs showed that a strict quality control at
all production stages is essential to ensure excellent performance
of the detector. An extensive list of quality control procedures was
established during the prototype phase and fully implemented
throughout the production phase. 

The tubes were produced in Italy by Pol.Hi.Tech and then shipped to 
the US. At Princeton University and Ohio State University they were assembled
into larger units and finally shipped to SLAC.

After production and before their final assembly into units
each cell of a tube is scanned with a radioactive source. For a good
tube the current is below 1\microA, with six dips in the current,
where the positions correspond to the wire holders. 
A typical failure is the occurrence of a continuous discharge, where
the current increases by at least 1\microA. In some cases the
discharge is self sustained and does not stop when the source is removed
from the cell. Tubes failing this test are opened in a clean room,
cleaned up and assembled again. If they fail the source scan after a
repair, the tube will be rejected. Failure modes are flakes of
graphite paint in the cell or impurities on the wire.

After transportation 
the tubes are visually checked for mechanical defects and tested for 
transportation damage by measuring the resistance and capacitance of
each HV channel.
All tubes are tested for gas leaks, where a half life time of a few
hours is necessary. Tubes which failed this test are manually searched
for leaks and usually repaired with Epoxy.

Each tube is HV conditioned after each production stage and after
arriving at SLAC. The HV is increased in steps of 200\V from 4900\V to
5900\V. A step is successfully completed when the current of the tube
is below 200\nA for at least 2 minutes. At 5900\V the current limit is
increased to 500\nA and the time limit to 10 minutes. For the finial
burn-in process the voltage is raised to 6000\V with the same current
and time limit as at 5900\V. Afterwards the tube is kept at 5900\V for
at least 10 hours. \fig~\ref{fig:HVcond} shows for two different tubes the
applied voltage and measured currents during the HV conditioning
process. \fig~\ref{fig:HVcond:good} shows the behaviour of an excellent
tube. The current increased only minimally with voltage increase and
the current is stable and $\sim$100\nA for the long-term part of the
process. \fig~\ref{fig:HVcond:bad} shows the burn-in
process. The current increases significant for the two highest
voltage steps, and then decreases slowly over time. After 10
hours the current is stable and below 100\nA as in the case of an
excellent tube. As long as the tube is kept on gas this behaviour will
not change.
The HV conditioning is repeated if the tube fails
the process at any HV step. If a tube still fails the procedure after
a few tries, it is rejected. Usually failed tubes develop a
self-sustained discharge with a current well above 1\microA.

\begin{figure}[p]
  \centerline{%
    \subfigure[Excellent tube.]{\includegraphics[width=0.9\textwidth]{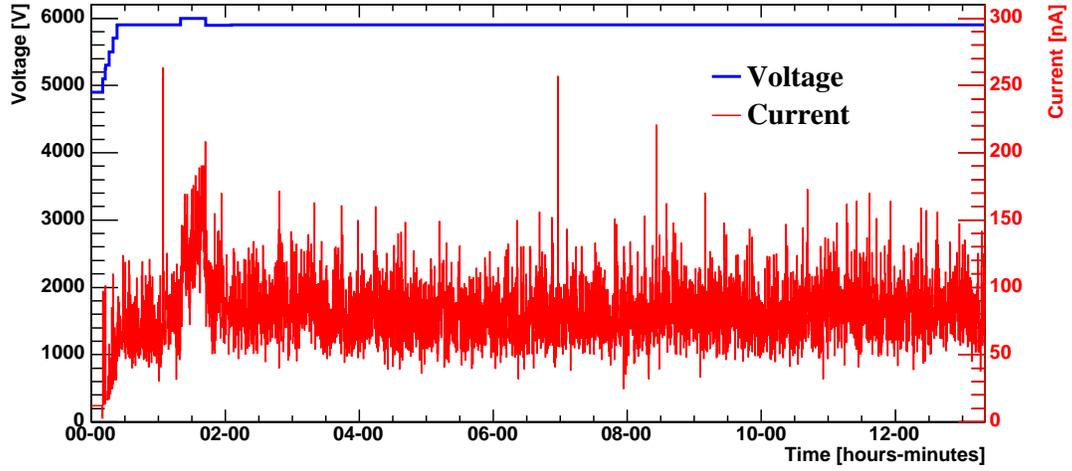}
      \label{fig:HVcond:good}}
  }
  \centerline{%
    \subfigure[Good tube after HV conditioning.]{\includegraphics[width=0.9\textwidth]{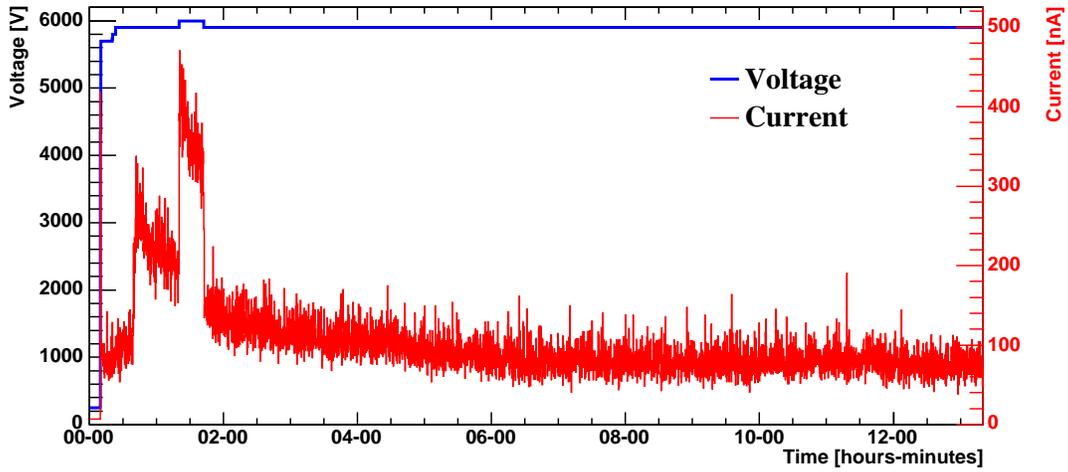}
      \label{fig:HVcond:bad}}%
  }
  \caption{For the HV conditioning process the voltage (blue) and 
    the current (red) is shown as a function of time.}
  \label{fig:HVcond}
\end{figure}

Finally the quality of each HV channel is checked by taking so called
single rates. This is the counting rate of cosmic muons as a
function of the HV. The HV is varied from 4900\V to 5900\V in steps of
100\V with a counting time of 100~seconds. \fig~\ref{fig:singlerates}
shows an example of the single rates for a good tube.
The counting rate starts quite low and then increases exponentially
around 5000\V. At $5100 - 5200\V$ the counting rate reaches a plateau,
which should be at least a few 100\V. In the case of an excellent tube
the plateau can go up to 5900\V.
A long and flat plateau is characteristic of a good tube.
\begin{figure}[p]
  \centering
  \includegraphics[width=0.9\textwidth]{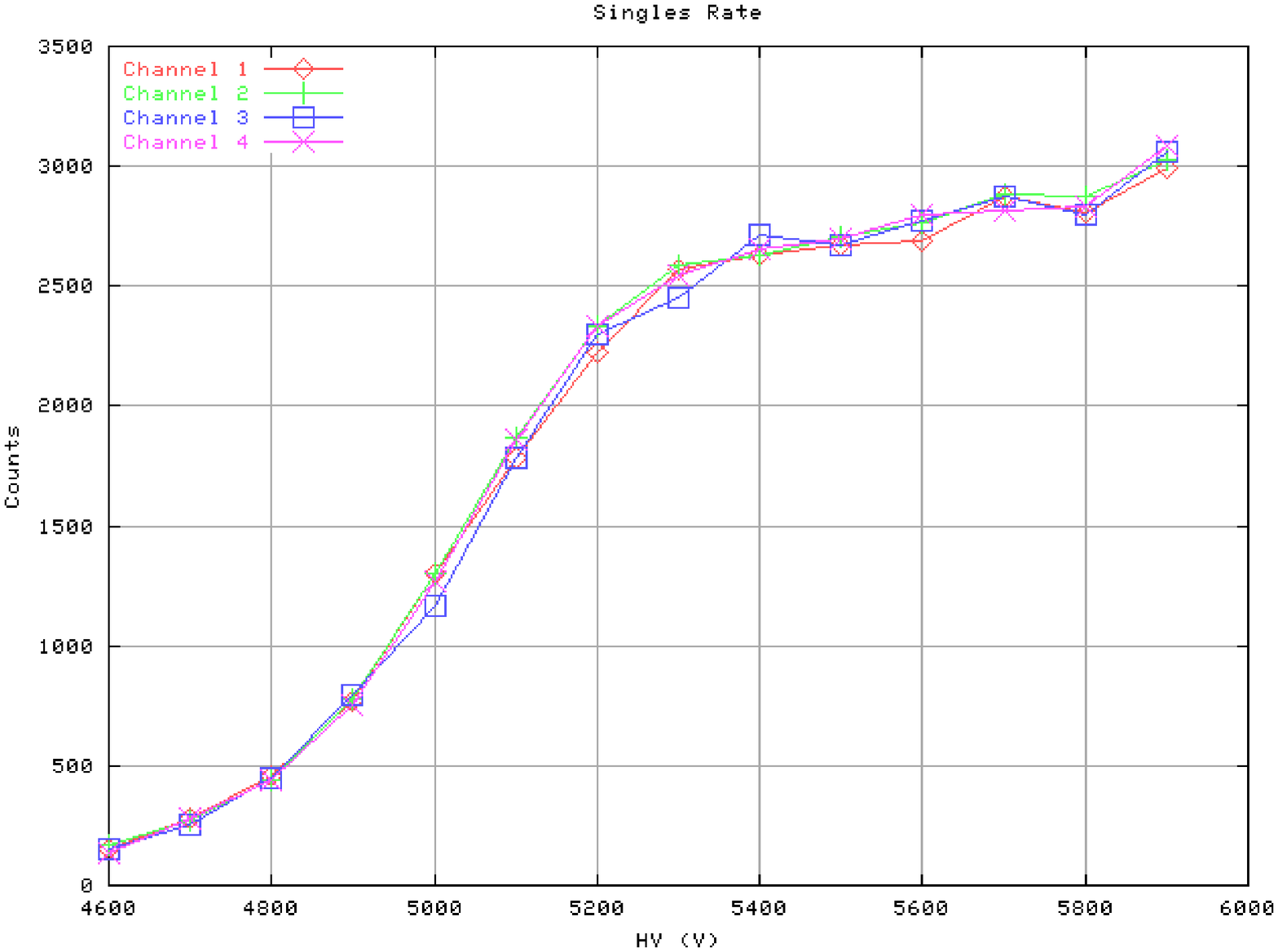}
  \caption{The counting rate using cosmic muons as a function of the
  HV is shown for the four signal channels in one tube.}
  \label{fig:singlerates}
\end{figure}

At the moment, the remaining $\sim$1300 tubes for the second
installation phase are under long-term testing at SLAC. They are kept
under gas flow and at the operating HV of 5500\V.  If the current of a
tube stays for 3 minutes above 500\nA, the HV is automatically ramped
down to 0\V.  The number of trips is recorded for later use in the
final selection process and tubes exhibiting a continuous
self-sustained discharge are disconnected from HV and rejected.

\subsection{Z-Planes}

The z-planes are 4\m long and up to 3.8\m wide, depending on the layer
in the IFR.
A plane consists of 96 copper strips, 35\mm wide with a spacing of
2\mm. They are glued on a copper ground plate, separated by Mylar. The
complete plane is vacuum laminated with a Mylar foil. 
The readout of 16 strips is grouped together into one flat cable,
where the strip and the cable are connected by a solder joint.

The design of the z-planes is very robust, nevertheless the planes
have to be tested to ensure their proper functioning. In addition
careful monitoring during installation is essential, when the planes 
may come under mechanical stress, 
in order to carry out repairs before the z-planes are inaccessible.
For monitoring the capacitance between ground and each readout channel
is measured. It should be around $\sim$5\nF. Very small or zero
capacitance is a clear indication of a broken solder joint. Only a few
broken solder joints have been found and most of them have been
repaired well before installation or while installing them.

\subsection{Gas System}

The gas used for the tubes is a (89:3:8) mixture of CO$_2$, Argon
and iso-butane. 
It is non flammable and has good quenching properties.
The gas mixing system is custom built. The mixing unit uses standard
technology and is based on mass flow controllers. 

Half of the tubes per layer are connected in sequence to one gas line.
The gas flow is monitored at the outlet with digital
bubblers\cite{Foulkes:2004qc}. The total flow rate of the system is
2.5$\,\mbox{l/min}$, which correspond to $\sim$1 volume changes per
day.

\subsection{High Voltage System}

The requirements of the HV system is high granularity and easy
accessibility. The segmentation of the system should be so good that
a single readout channel of a tube can be removed from the supply so
that the rest of the tube can be operated without any problems.  On
the other hand, HV control of a whole layer and current monitoring of a
single tube is sufficient.

The design of the HV power supplies was guided by these requirements.
They are custom built~\cite{gabriele} and can operate between 0 and
6000\V. Each supply consists of 80 HV channels in four
independent HV groups.  The current monitoring, current limits and
trip times are on a channel by channel basis. Each channel is equipped
with a hardwired over-current protection circuit based on a design
from the ZEUS muon system.
Starting around 3\microA the effective voltage is lowered, depending
on the drawn current.

Each HV channel is split into four pins allowing access to a quarter of
a tube. If a readout channel of a tube evolves a recurring problem
over a long period of time, this channel can be easily removed from
the HV supply and the other three quarters of the tube can be operated
without any problems.

The HV cables are built from multi-conductor Kerpen cable and consist
of two parts connected with a custom-inline connector: a short-haul
cable mounted on the tube and a long-haul cable going
from the detector to the power supplies.

\subsection{Front End Electronics}

The front end electronics were specially developed for the needs of the
LSTs. They interface to the existing RPC-FIFO and are then read out
into the standard \babar DAQ system. On each motherboard four daughter boards
can be installed. The signal from the wires and the strips are
different in polarity and shape and this difference is taken care of on
the daughter board level. Common features are implemented on the
motherboard. A daughter board has 16 analog input channels. The signals
are amplified, discriminated with an adjustable threshold and
converted to 1-bit digital hit signals . No other information is kept,
e.~g. timing, charge and other shaping information are not passed on to
the DAQ system.

The crates with the front end electronics are located in the near
vicinity of the \babar detector inside the shielding wall. They are
not mounted directly on the detector to give better
accessibility.

\section{First Installation Phase}

The first installation phase was scheduled for August to October 2004.
The RPCs from the inner 18 layers of the top and bottom
sextant were removed. In 12 of the 18 layers z-planes and LSTs were
installed. In the remaining 6 layers brass was installed to increase
the total absorption length 
and compensate for the loss in absorption length between the 18th and
19th layer. In total 24 z-planes for a total of 2284 z strips and 388
tubes were installed, which results in 1522 $\phi$ readout channels. 
188 HV cables were used and connected to 6 power
supplies. One additional power supply was installed in the case it is
needed for problematic tubes which need to be isolated. 332 signal
cables were used and connected to 84 FEC boards.

\section{Second Installation Phase}

The second installation phase is scheduled for autumn 2006. The
remaining 4 sextants of RPCs will be replaced with LSTs. This will be
776 tubes and 48 z-planes. Additionally, 14 HV power supplies and
168 FEC cards will be installed.

\section{Operations}

The LSTs have been maintained at the operational voltage of 5500\V
since October 2004. After an extended shutdown, \babar\
resumed data taking in March 2005 and collected until October 2005
$\sim$60\invfb of \Pep\Pem\ collisions at the \PgUc\ resonance. This
dataset will be increased to $\sim$250\invfb by summer 2006,
when the next shutdown and the second installation phase is
scheduled. This dataset will be the same size as the one collected
from the begin of \babar\ up to summer 2004 but with two
sextants of excellent working muon detectors.

The occupancy of each wire channel
is constantly monitored online during data taking.  The plateau of
each cell is measured every month and all channels have a good plateau
except for 5 channels. These channels trip often. The problems are
diagnosed to be located between the power supplies and the cell wires.
They are isolated in the extra HV supply. At the moment they are
operated a few 100\V below the nominal operation voltage. In summary
more than 99.6\% of all channels are working perfectly with no
decrease over time visible.

The z strip occupancy is also monitored online. All channels except 5
give good readings. These dead channels have been tracked down to
broken solder joints and the number is constant over the time. In
summary more than 99.7\% of all channels are working perfectly.

The effect of these dead or not properly working channels on the
physics performance is expected to be negligible because of the high
granularity in wires, strips and layers.

For every run the efficiency per layer is determined from a radiative
di-muon sample. The average efficiency is above 90\%, consistent with 
the geometrical acceptance of the LST. The efficiency is 
constant over time.

\fig~\ref{fig:muonselector} shows the pion rejection rate as a
function of the muon efficiency for high energy muons
($2\GeV<p<4\GeV$) using a neural network based muon selection
algorithm for the years 2000, 2004 and 2005. The year 2005 data is
split up into LSTs and RPCs.  The decrease in muon
efficiency and pion rejection from 2000 to 2005 is clearly visible. In
2005 the maximal muon efficiency using RPCs is 88\% with poor pion
rejection compared to almost 94\% muon efficiency with moderate pion
rejection in 2000.
With the LSTs the overall performance is even better than the initial
performance of the RPCs in the first year of operation (2000). For a
fixed muon efficiency, the LSTs always give a higher pion rejection
rate. The muon efficiency can reach almost 94\% with a moderate pion
fake rate. 
This is a very clear indication of the success of the muon system
upgrade with LSTs.
\begin{figure}[p]
  \centering
  \includegraphics[width=0.9\textwidth]{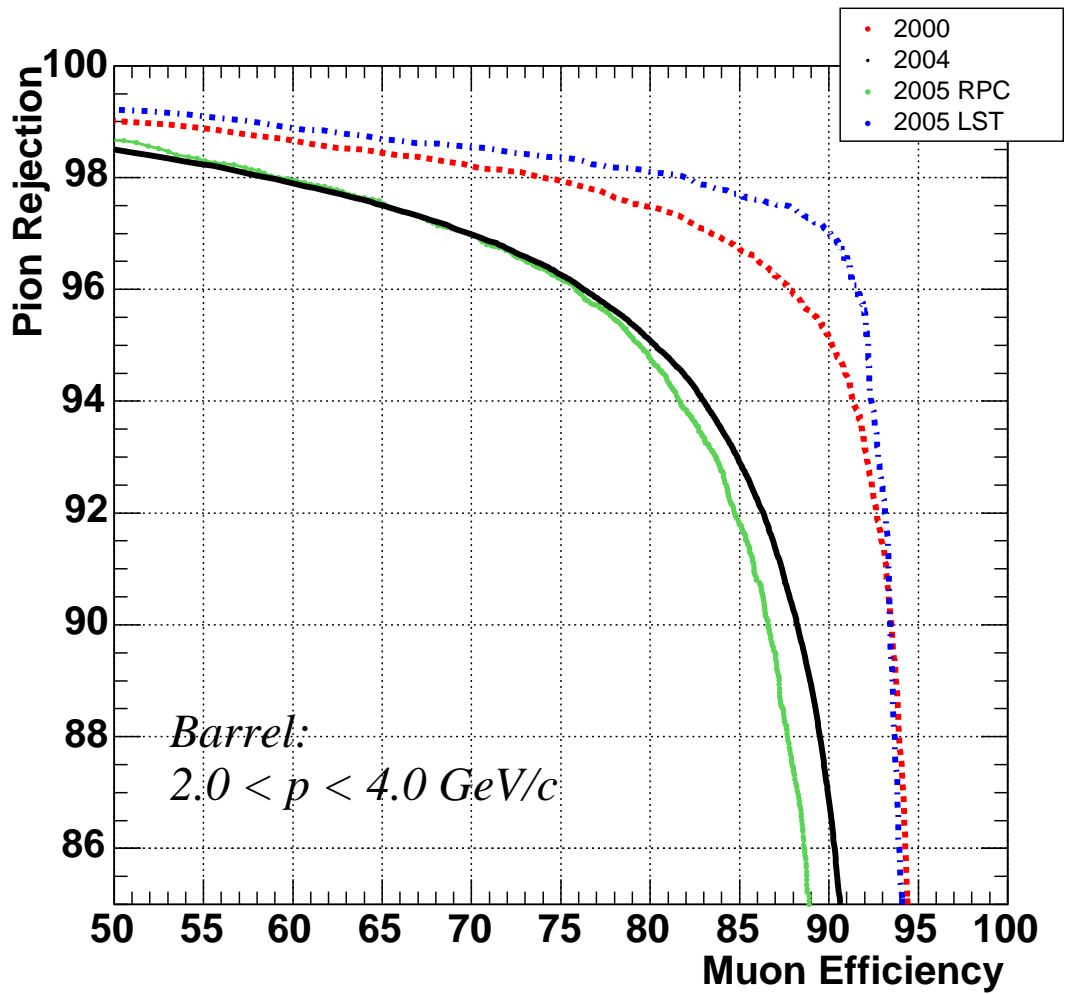}
  \caption{The pion rejection rate as a function of the muon
  efficiency is shown using a muon selector based on a neural network
  for high energy muon.
  The data for the RPCs in the years 2000, 2004 and 2005 is plotted in
  red, black and green, respectively. The data for the LSTs in the
  year 2005 is shown in blue.}
  \label{fig:muonselector}
\end{figure}

\section{Conclusion}
The summer 2004 installation was very successful. Almost 100\% of the
$\phi$-readout and z-readout channels are working. The efficiency per
layer is consistent with the geometrical acceptance. This excellent
per-layer efficiency translates into an excellent muon selection
efficiency of up to 94\% with a moderate pion fake rate. The overall
performance with LSTs is even better than the initial performance of
the RPCs. This excellent performance is due to the stringent quality
control of all detector components at all production stages.

\appendix

\section*{The \babar-LST Group}

The following institutes form the \babar-LST group:
Universit\`a di Ferrara and INFN,
Laboratori Nazionali di Frascati dell'INFN,
Universit\`a di Genova and INFN,
Lawrence Livermore National Laboratory,
Massachusetts Institute of Technology,
Ohio State University,
Universit\`a di Padova and INFN,
Universit\`a di Perugia and INFN,
Queen Mary, University of London,
Princeton University,
University of British Columbia,
University of  California at Santa Babara,
University of Colorado,
Colorado State University,
Universit\`a di Torino and INFN,
University of Oregon,
Universit\`a di Roma La Sapienza and INFN, 
University of California at San Diego,
and
Stanford Linear Accelerator Center.

\section*{Acknowledgment}
The authors would like to thank the Stanford Linear Accelerator Center
and the \babar-LST group for the kind hospitality.
This work is supported by the
US Department of Energy,
the Istituto Nazionale di Fisica Nucleare (Italy),
and the Particle Physics and Astronomy Research Council (United Kingdom).



\begin{thebibliography}{1}
\bibitem{Aubert:2001tu}
  B.~Aubert {\it et al.}  [BABAR Collaboration],
  Nucl.\ Instrum.\ Meth.\ A {\bf 479}, 1 (2002)
  [arXiv:hep-ex/0105044].
\bibitem{Piccolo:2001zy}
  D.~Piccolo {\it et al.},
  Nucl.\ Instrum.\ Meth.\ A {\bf 515}, 322 (2003).
\bibitem{Anulli:2003bk}
  F.~Anulli {\it et al.},
  Nucl.\ Instrum.\ Meth.\ A {\bf 508}, 128 (2003).
\bibitem{Anulli:2004zu}
  F.~Anulli {\it et al.},
  Nucl.\ Instrum.\ Meth.\ A {\bf 539}, 155 (2005).
\bibitem{Foulkes:2004qc}
  S.~Foulkes {\it et al.},
  Nucl.\ Instrum.\ Meth.\ A {\bf 538}, 801 (2005).
\bibitem{gabriele}
  G.~Benelli {\it et al.}
  ``The \babar LST Detector High Voltage System: Design and Implementation,''
  these proceedings.
\end{thebibliography}
\end{document}